\documentclass[10pt,aps,prl,reprint,amsmath,showpacs,preprintnumbers,twocolumn,amssymb,superscriptaddress,longbibliography]{revtex4-1}
\usepackage{amsmath}
\usepackage{graphicx}
\usepackage{bm}
\usepackage{mathptmx, textcomp}
\usepackage{tabularx}
\usepackage{siunitx}
\usepackage{color}
\definecolor{bl}{rgb}{0, .1, .6}
\definecolor{rd}{rgb}{1,0,.2}

\usepackage[colorlinks=true, citecolor = bl, linkcolor = bl, urlcolor=bl, pdfborder={0 0 0}]{hyperref}
\usepackage{natbib}
\usepackage{tabularx}
\usepackage{lipsum}
\newcommand{\mum}{\SI{}{\micro\meter}}

\newcommand{\Hz}{{\rm Hz}}
\newcommand{\g}{{\rm G}}
\newcommand{\mg}{{\rm mG}}
\newcommand{\ms}{{\rm ms}}
\newcommand{\be}{\begin{eqnarray}}
\newcommand{\ee}{\end{eqnarray}}
\newcommand{\abg}{a_{\rm bg}}

\newcommand*{\vcenteredhbox}[1]{\begingroup
\setbox0=\hbox{#1}\parbox{\wd0}{\box0}\endgroup}
\begin{document}

\title{Scissors mode of dipolar quantum droplets of dysprosium atoms}

%\author{Igor Ferrier-Barbut, Matthias Wenzel, Fabian B\"ottcher, Tim Langen, Tilman Pfau}
%\affiliation{5. Physikalisches Institut and Center for Integrated Quantum Science and Technology,
%Universit\"at Stuttgart, Pfaffenwaldring 57, 70550 Stuttgart, Germany}

\author{Igor Ferrier-Barbut}
\email{i.ferrier-barbut@physik.uni-stuttgart.de}
\affiliation{5. Physikalisches Institut and Center for Integrated Quantum Science and Technology IQST,
Universit\"at Stuttgart, Pfaffenwaldring 57, 70550 Stuttgart, Germany}
\author{Matthias Wenzel}
\affiliation{5. Physikalisches Institut and Center for Integrated Quantum Science and Technology IQST,
Universit\"at Stuttgart, Pfaffenwaldring 57, 70550 Stuttgart, Germany}
\author{Fabian B\"ottcher}
\affiliation{5. Physikalisches Institut and Center for Integrated Quantum Science and Technology IQST,
Universit\"at Stuttgart, Pfaffenwaldring 57, 70550 Stuttgart, Germany}
\author{Tim Langen}
\affiliation{5. Physikalisches Institut and Center for Integrated Quantum Science and Technology IQST,
Universit\"at Stuttgart, Pfaffenwaldring 57, 70550 Stuttgart, Germany}
\author{Mathieu Isoard}
\altaffiliation{Present address: LPTMS, CNRS, Univ. Paris-Sud, Universit\'e Paris-Saclay, 91405 Orsay, France}
\affiliation{INO-CNR BEC Center and Dipartimento di Fisica, Universit\`a di Trento, 38123 Povo, Italy}
\author{Sandro Stringari}
\affiliation{INO-CNR BEC Center and Dipartimento di Fisica, Universit\`a di Trento, 38123 Povo, Italy}
\author{Tilman Pfau}
\affiliation{5. Physikalisches Institut and Center for Integrated Quantum Science and Technology IQST,
Universit\"at Stuttgart, Pfaffenwaldring 57, 70550 Stuttgart, Germany}

\begin{abstract}
We report on the observation of the scissors mode of a single dipolar quantum droplet. The existence of this mode is due to the breaking of the rotational symmetry by the dipole-dipole interaction, which is fixed along an external homogeneous magnetic field. By modulating the orientation of this magnetic field, we introduce a new spectroscopic technique for studying dipolar quantum droplets. This provides a precise probe for interactions in the system allowing to extract a background scattering length for \textsuperscript{164}Dy of $69(4)\,a_0$. Our results establish an analogy between quantum droplets and atomic nuclei, where the existence of the scissors mode is also only due to internal interactions. They further open the possibility to explore physics beyond the available theoretical models for strongly-dipolar quantum gases.
\end{abstract}

\maketitle

The recent observation of quantum droplets in dipolar Bose-Einstein condensates (dBEC) \cite{Kadau:2016,FerrierBarbut:2016,Chomaz:2016} and in BEC mixtures \cite{Petrov:2015,Cabrera:2017,Semeghini:2017} opens the opportunity to bridge the gap between dense quantum liquids such as atomic nuclei and helium, and very dilute ultracold atomic samples. This link was reinforced by the observation of the self-bound character of quantum droplets \cite{Schmitt:2016,Cabrera:2017,Semeghini:2017}. Prior to this, several phenomena shared by dense quantum liquids and dilute superfluids were observed. In particular, the so-called scissors mode first observed in nuclei \cite{LoIudice:1978,Lipparini:1983,Enders:1999} was later predicted and observed in Bose-Einstein condensates in anisotropic external potentials \cite{GueryOdelin:1999,Marago:2000}. In nuclei this mode corresponds to the out-of-phase rotation of the neutrons and protons, and in BECs it is an angular oscillation around the anisotropy axis \cite{Pitaevskii:2016}. Its existence is a marker of the breaking of a rotational symmetry. A stark difference however between BECs and nuclei is that in the latter, the scissors mode arises only due to internal interactions. In contact interacting BECs this mode exists only in an anisotropic external potential, it vanishes in cylindrically-symmetric traps.\par
 Quantum droplets are liquid-like objects, bound by a mean-field attraction and stabilized by beyond mean-field effects \cite{Petrov:2015,FerrierBarbut:2016}. Their collective modes are a revealing probe for their internal properties, \cite{Chomaz:2016,Wachtler:2016,Baillie:2017}. The scissors mode was theoretically explored in the context of dBEC in \cite{vanBijnen:2010}. Here, we demonstrate that the anisotropy of the dipole-dipole interaction (DDI), set by the external homogeneous magnetic field, leads to a well-defined scissors mode in dipolar quantum droplets even in cylindrically-symmetric trapping geometries. We parametrically excite this mode, and the high frequencies observed reveal the very strong intrinsic anisotropy of dipolar quantum droplets. In addition, it is known that this mode is well-defined only for low excitation amplitude, while it is non-linearly coupled to other low-frequency modes for large excitation angles \cite{Pitaevskii:2016}. We observe clear signatures of this non-linear mode coupling and use such coupling to excite a low-frequency mode. Altogether these measurements represent a strong test of internal interactions in the droplets, we thus extract the value of the s-wave background scattering length of \textsuperscript{164}Dy with good precision. We put this in perspective with previous measurements, highlighting the two- and many-body physics at play in dipolar quantum droplets of dysprosium.\par
 \begin{figure*}[hbtp]
\vcenteredhbox{\includegraphics[width=.5\columnwidth]{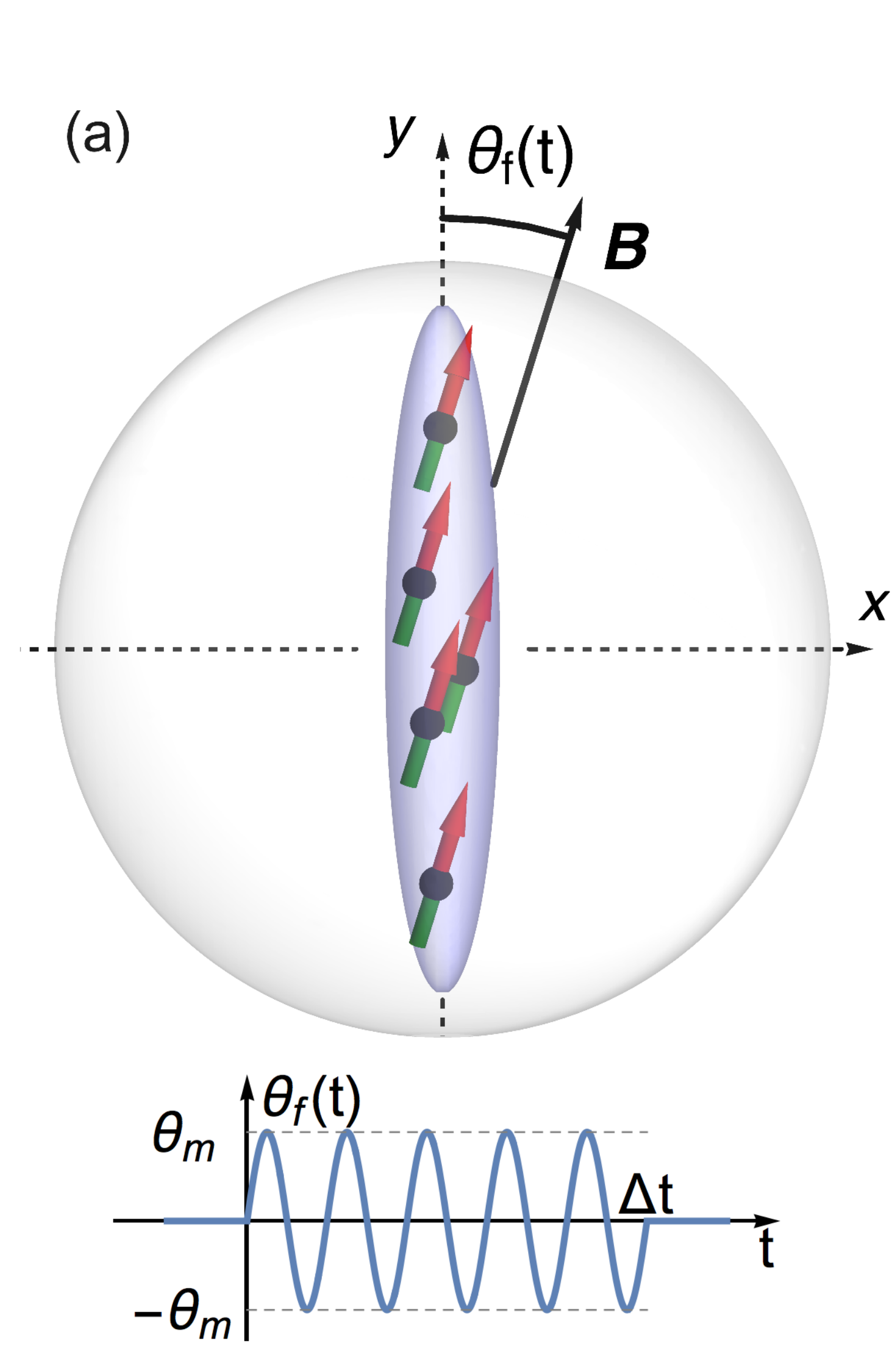}}
\vcenteredhbox{\includegraphics[width=1.4\columnwidth]{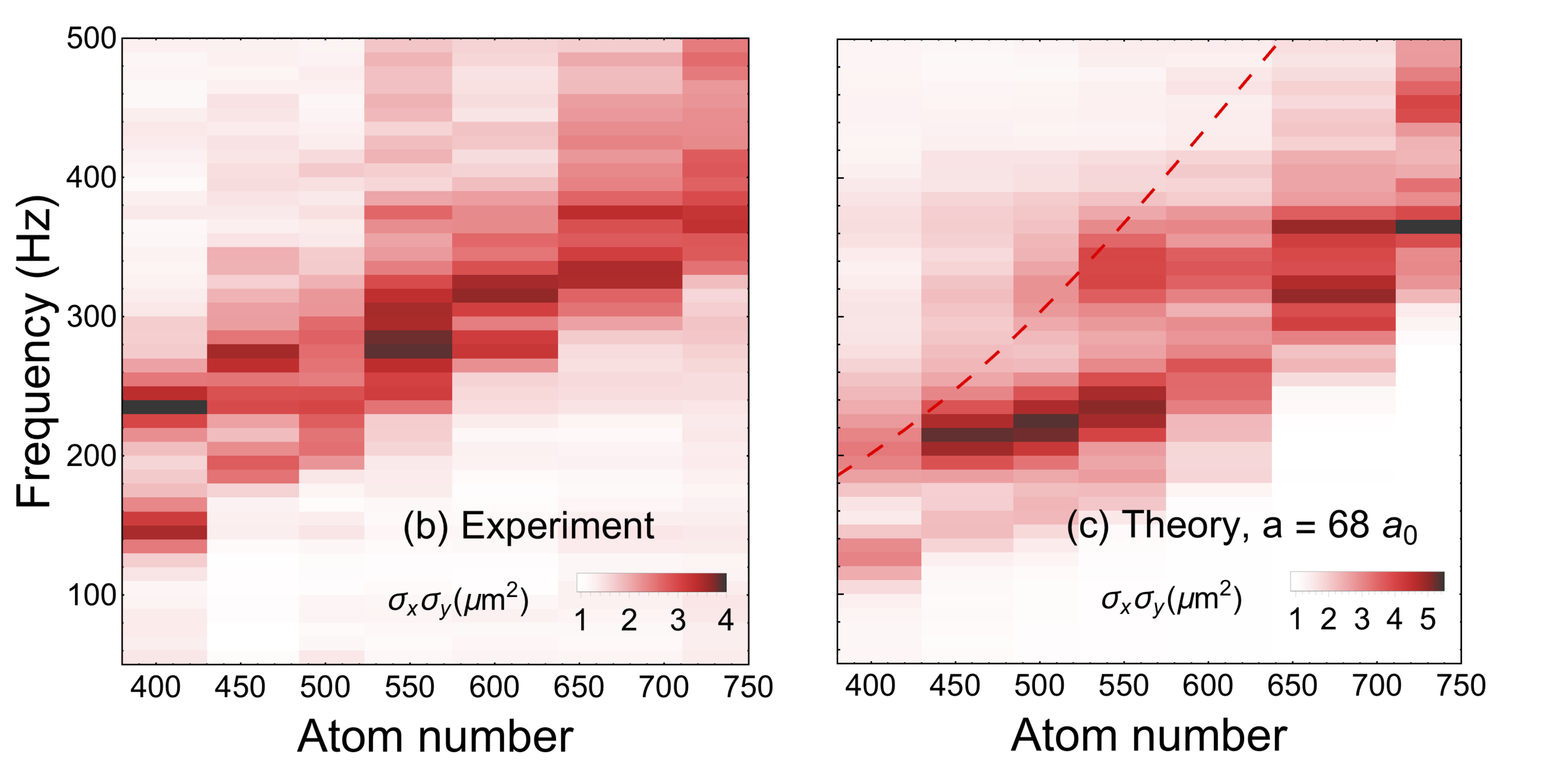}}
\caption{(a) Experimental method: The quantum droplet is held in a cylindrically symmetric trap (around $z$), the orientation of the field is modulated around its mean value along $y$ for $\Delta t=20\,\ms$ at variable frequency. The amplitude $\theta_{\rm m}$ follows: $\theta_{\rm m}(f)=12^\circ/\sqrt{f/100\,\Hz}$. (b): Experimental response measured as a growth in the visible size of the droplet $\sigma_x\times\sigma_y$ as a function of atom number and modulation frequency. (c) The theory shows the same value extracted from solutions of the equations of motion within the Gaussian ansatz using the best fit value of $a=68\,a_0$. It takes into account the finite excitation time as well as  departure from the linear response regime. The shot-to-shot fluctuations in atom number are simulated and finite resolution is also implemented in the theory calculation of $\sigma_x\times\sigma_y$. Note the different scales for theory and experiments, the $\sim 20\,\%$ difference in droplet size could be due to imaging miscalibration. The dashed line shows the theoretical scissors mode frequency in linear response theory, see \cite{SupMat}.}
\label{Fig:ModulationResponse}
\end{figure*}

\textit{Theory} The scissors mode, corresponding to an angular oscillation, is naturally excited by the $z$-component of the  angular momentum operator $\hat{L_z} = \sum_{k=1}^N(x_k\, p^y_k - y_k \, p^x_k)$. In the experiment, this corresponds to a rotation of the external magnetic field axis around $\hat z$. We consider here dipoles oriented along the $y$ direction (see Fig.~\ref{Fig:ModulationResponse} (a)), thereby breaking rotational invariance in the $xy$-plane, even in the presence of a cylindrically-symmetric trap. Employing linear response theory one can derive a rigorous upper bound to the frequency of the scissors mode in the form \cite{Pitaevskii:2016}
\begin{equation}
\hbar \omega_{\rm sc} = \sqrt{\frac{m_1}{m_{-1}}},
\label{ratio_sm}
\end{equation}
where $m_1 = \hbar^2 \int{d\omega \, \omega \,S_{L_z}(\omega)}$ is the energy-weighted moment of the dynamical structure factor $S_{L_z}(\omega)$, relative to the angular momentum operator, while $m_{-1}=\int{d\omega/\omega \,S_{L_z}(\omega)}$ is the inverse energy-weighted moment. Both moments $m_1$ and $m_{-1}$ incapsulate important physical information on the  scissors mode. The $m_1$ moment can be in fact expressed in terms of a double commutator involving the Hamiltonian of the system as
\begin{equation}
m_1(\hat{L}_z)  = \frac{1}{2}\left<[ \hat{L}_z,[ \hat{H}, \hat{L}_z ] ]\right>
\label{m1}
\end{equation}
and can be regarded as an effective restoring force parameter for the scissors oscillation. Here
$\left< \,.\,\right>$ is the average taken on the  equilibrium configuration of the system. The non-vanishing of the commutator $[\hat{H}, \hat{L}_z ]$ is the consequence of the breaking of rotational invariance. This can be due either to the presence of an anisotropic trapping potential, and/or to the presence of the dipolar interaction. In the case we are interested in, of isotropic harmonic trapping ($\omega_x=\omega_y\equiv \omega_\perp$), (or in the absence of trapping), only the dipolar interaction contributes to the commutator and the $m_1$ sum rule takes the useful form (see Supplemental Material \cite{SupMat}\nocite{Glaum:2007,Giovanazzi:2006}):
\begin{equation}
m_1 = \frac{\hbar^2}{2} \left( \langle V_{\rm dd}^x\rangle - \langle V_{\rm dd}^y\rangle \right) \; ,
\label{m1dipolar}
\end{equation}
where 
$\langle V_{\rm dd}^\alpha\rangle=\int d\bm{r}d\bm{r'}n(\bm{r})V_{\rm dd}^\alpha(\bm{r}-\bm{r'})n(\bm{r'})$
and
$V_{\rm dd}^\alpha(\bm r)=\frac{\mu_0\mu^2} {4\pi r^3}\left(1-3\frac{\alpha^2}{r^2} \right)$, $\alpha = x,\;y,\;z$. Equation (\ref{m1dipolar}) emphasizes the crucial role played by the dipolar interaction which causes an asymmetry between the ground state expectation values $\langle V_{\rm dd}^x\rangle$ and $\langle V_{\rm dd}^y\rangle$. For  \textsuperscript{164}Dy $\mu \approx 10\,\mu_{\rm B}$ with $\mu_{\rm B}$ the Bohr magneton. This defines the dipolar length $a_{\rm dd} = \frac{\mu_0\mu^2m}{12\pi \hbar^2}$, compared to the $s$-wave scattering length $a$ via $\varepsilon_{\rm dd}=a_{\rm dd}/a$.\par
Differently from $m_1$, the inverse energy-weighted moment  $m_{-1}$ cannot be written in terms of commutators, but can be usefully identified in terms of the moment of inertia $\Theta$ of the system. Actually the moment $m_{-1}$ corresponds, apart from a factor $1/2$, to the static response of the system to an angular momentum perturbation of the form $-\omega \hat{L}_z$. The moment of inertia, which provides the mass parameter of the scissors oscillation,  is very sensitive to superfluidity and for a Bose-Einstein condensate at zero temperature is given by the expression \cite{Pitaevskii:2016}
\begin{equation}
\Theta = 2\,m_{-1} = m\,\frac{(\langle y^2\rangle -\langle x^2\rangle)^2}{\langle y^2\rangle +\langle x^2\rangle}
\label{inertia}
\end{equation} 
which follows from the irrotationality constraint characterizing the superfluid velocity. For an axi-symmetric configuration the moment of inertia of a superfluid then identically vanishes.\par
All the average quantities characterizing the moments $m_1$ and $m_{-1}$  can be evaluated using the Gaussian ansatz  
%\begin{equation}
$	\psi(\boldsymbol{r})=\frac{\sqrt N}{\pi^{3/4}\bar \sigma^{3/2}}e^{\sum_\alpha -\frac{\alpha^2}{2\,\sigma_\alpha^2}}$
%\label{Eq:psi}
%\end{equation} 

for the order parameter relative to the ground state, with $\alpha \in\{x,y,z\}$ and $\bar\sigma^3=\sigma_x\sigma_y\sigma_z$. The square radii, entering the expression for the moment of inertia, are given by  $\langle r_i^2\rangle=\sigma_i^2/2 $. The values $\langle V_{\rm dd}^x\rangle$ and $\langle V_{\rm dd}^y\rangle$, entering the $m_{1}$ sum rule, can be also calculated using the Gaussian ansatz and in the general case of anisotropic trapping we recover the results for dipolar BECs obtained in \cite{vanBijnen:2010}. The expression for the resulting scissors mode frequency $\hbar \omega_{\rm sc} =	\sqrt{\frac{m_1}{m_{-1}}}$ is reported as Eq. S13 in \cite{SupMat}. It is worth noticing that  the calculated equilibrium sizes $\sigma_i$, obtained through a variational  procedure applied to the energy of the system,  strongly depend on the scattering length $a$ \cite{Wachtler:2016,Bisset:2016} which gives an implicit dependence  of the scissors frequency on $a$. We have checked that a time-dependent simulation, based on extended Gross-Pitaevskii (eGPE) theory, agrees in the linear limit  with the sum rule  value for the scissors frequency calculated using the Gaussian ansatz. The basic ingredients underlying  the dynamics of the scissors mode are indeed well captured by the sum rule approach. Actually  our result takes into account both the breaking of rotational symmetry caused by the dipolar interaction [see Eq.~(\ref{m1dipolar})] and  the superfluid expression (\ref{inertia}) for the moment of inertia which,  however, does not  differ significantly from the classical rigid value $m( \langle x^2\rangle+\langle y^2\rangle)$  due to the large anisotropy of the dipolar droplet characterizing the present experimental conditions. The sum rule approach provides a reliable estimate of the scissors frequency in the linear regime. Experiments involve however relatively large amplitudes of the oscillation and for a systematic quantitative comparison it is  useful to develop a time-dependent extension of the variational approach. It is based on the Lagrangian formalism, where the Gaussian ansatz is generalized to include a phase playing the role of a velocity potential as already employed for dipolar BECs in \cite{Yi:2001} and in \cite{Wachtler:2016} for quantum droplets, now with an additional parameter accounting for the orientation $\theta$ of the droplet in the $xy$-plane. The new calculation, which corresponds to solving the equations of motion for the four degrees of freedom $\sigma_{x,y,z}$ and $\theta$,  accounts for the excitation of the scissors mode as well as of additional low frequency modes of quadrupole and compression nature which play an important role in the non-linear limit as we will discuss below.\par

\textit{Experiments} We perform experiments on dipolar quantum droplets in an optical dipole trap, in which we obtain lifetimes of several hundreds of milliseconds. The trapping configuration was presented in \cite{Wenzel:2017}. Here the trap has fixed frequencies of $f_x=f_y=40(1)\,\Hz,\;f_z=950\,\Hz$ ($\hat z$ being along gravity), thus isotropic in the $xy$-plane, as assumed by the theory presented above. The magnetic field is always oriented in this plane, initially along $\hat y$. In such geometry, quantum droplets can exist for smaller atom numbers than in free space and for very small atom numbers outside the range of our experiments, similarly to \cite{Cheiney:2017} they transform into solitons \cite{Tikhonenkov:2008,Koberle:2012} which differ from quantum droplets by being stable even without beyond mean-field corrections, see \cite{SupMat}. To extract their properties, we fit column-integrated images with a Gaussian distribution $\bar n=\frac{N}{\pi \sigma_x\sigma_y}\exp\left(-\frac{x^2}{\sigma_x^2}-\frac{y^2}{\sigma_y^2}\right)$. We obtain a typical size along $\hat y$ of $\sigma_y\sim1.5\,\mum$ while the extent along $\hat x$ is smaller than our resolution \cite{SupMat}. We create single droplets containing a few hundred atoms. Their density being initially high, the atom number decays fast from $N\approx750$ down to $N\approx400$ via three-body losses. The systematic uncertainty on the number of condensed atoms within the droplet is $\delta N/N=0.25$ \cite{SupMat}. The absolute value of the magnetic field is fixed to be $B_0=800\,\mg$, far from any Feshbach resonance \cite{Baumann:2014} so that the scattering length takes the low-field background value $\abg$.\par
In the first set of experiments we parametrically excite the scissors mode by adding an oscillating magnetic field along $\hat x$, exemplified in Fig.~\ref{Fig:ModulationResponse} (a). The $x$-field follows: $B_x(t)=B_{x0}\sin(2\pi f\,t)$, with a variable frequency $f$ and a maximum amplitude of $B_{x0}\leqslant200\,\mg$. The angle of the field with respect to the $y$ axis $\theta_{\rm f}$ is then $\theta_{\rm f}(t)\simeq B_x(t)/B_0=\theta_{\rm m}\,\sin(2\pi f\,t)$. The modulation time is $\Delta t =20\,\ms$, chosen so that atom numbers variations are small $\Delta N/ N\leqslant10\%$ during this time. Since $\Delta t$ is kept fixed, to keep a constant pulse `energy' we decrease the modulation amplitude with a $1/\sqrt{f}$ scaling: $\theta_{\rm m}(f)=\theta_0/\sqrt{f/100\,\Hz}$. We first perform our experiments with $\theta_0=12^\circ$.\par
%The absolute value of the field varies very little due to the modulation ($\frac{\delta |B|}{B_0}\leqslant4\times 10^{-2}$). Since we are far from Feshbach resonances, we expect that the varying absolute value of the field does not cause any excitation. We have verified this by modulating only the field magnitude at fixed orientation and have observed no response at all from the droplet \cite{SupMat}.\par
{When the field direction is modulated, we observe a clear excitation of the scissors mode. We show in \cite{SupMat} that this is not due to the very small modulation of the absolute field ($\frac{\delta |B|}{B_0}\leqslant4\times 10^{-2}$). The excitation is seen as an increase in the observed size of the droplet: $\sigma_x\times \sigma_y$. Since the atom number varies with time, we can investigate the variation of the response with atom number. We observe a clear dependence, shown in Fig.~\ref{Fig:ModulationResponse} (b). The maximum response frequency clearly increases with atom number. We note that several features are also visible in Fig.~\ref{Fig:ModulationResponse} (b). In particular a splitting into two lines is observable at low atom number. These characteristics signal that the observed response contains more than a simple parametric excitation of the scissors mode in the linear response regime. The rather large anisotropy of dipolar quantum droplets even in symmetric traps leads to a well defined scissors mode. However as we impose values of the excitation angle close to the deformation $\frac{\sigma_y^2-\sigma_x^2}{\sigma_x^2+\sigma_y^2}$ of the atomic cloud, we expect to approach the regime where the scissors mode is not well defined as it couples to other low-lying modes \cite{GueryOdelin:1999}.

\begin{figure}[hbtp]
\includegraphics[width=\columnwidth]{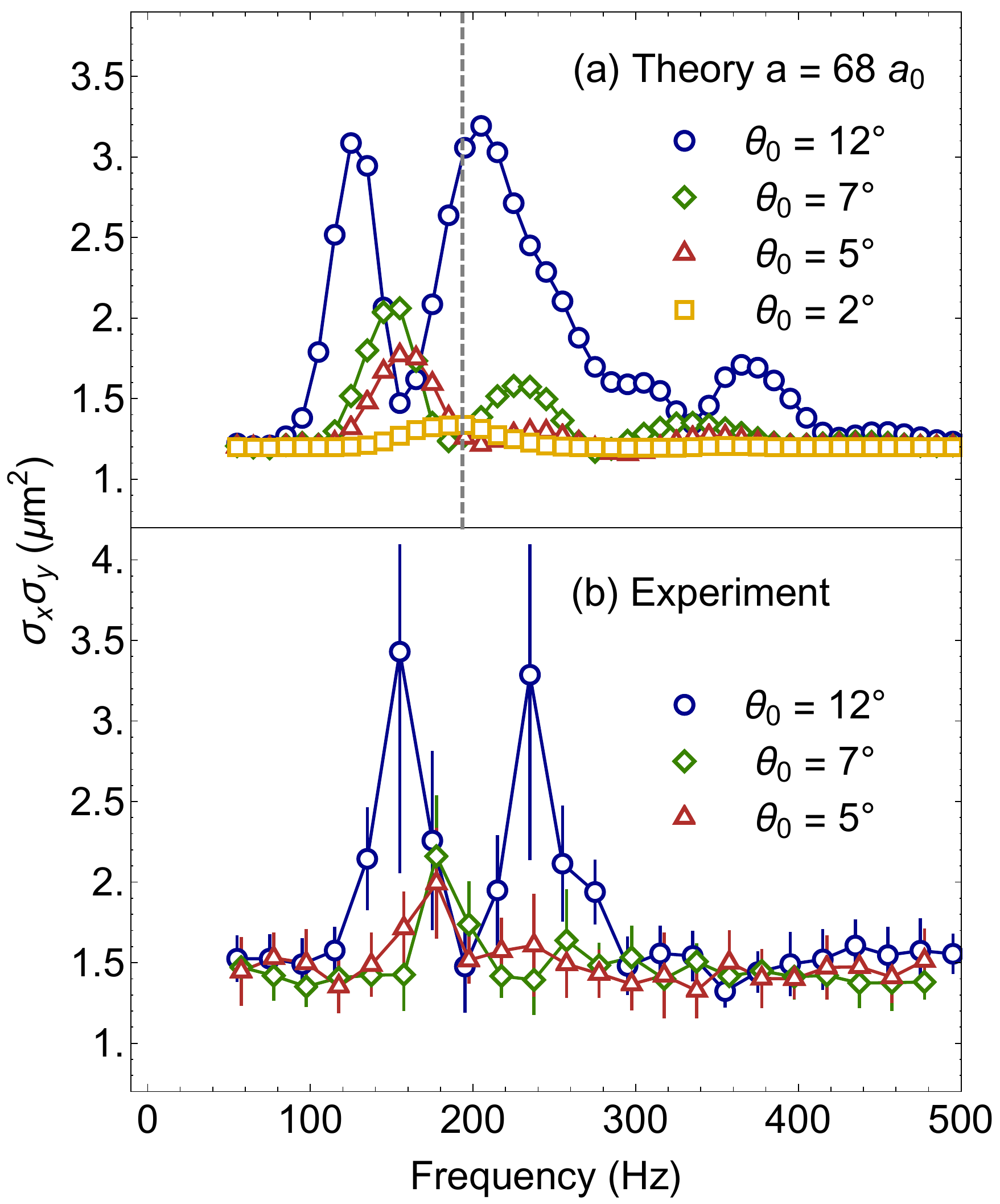}
\caption{Variation of the response with the angle modulation amplitude. Data and theory in Fig.~\ref{Fig:ModulationResponse} corresponds to $\theta_0=12^\circ$. (a) Theory. We take $a=68\,a_0$ as it gives the best fit over the whole range of atom number in Fig.~\ref{Fig:ModulationResponse}, this explains the slight horizontal disagreement with experiment. For increasing excitation amplitude, the response increases, and a line splitting is observed due to departure from the linear response regime. The dashed line shows the linear-repsonse prediction. (b) Experiment, with atom number $390(100)$. The same increase in response and line splitting is observed. We surmise that the small peaks at high frequency absent in the experiment are due to the simplifying assumptions of our model.}
\label{Fig:VaryingAmplitude}
\end{figure}

To confirm that the line splitting for the lowest atom numbers is due to hybridization of the scissors mode, we perform experiments at a fixed atom number $N=390(100)$, but for varying amplitude. This is represented in Fig.~\ref{Fig:VaryingAmplitude} (b), for low amplitude we obtain a much lower response, requiring much more data averaging to reach a sufficient signal to noise ratio. But we do observe that only one peak appears at lower amplitude, confirming that departure from the linear regime occurs for the amplitudes used in Fig.~\ref{Fig:ModulationResponse} (b). In order to capture these effects which come from a coupling between the different lowest-lying modes of the system, we compare the experimental results with the predictions of the  variational time-dependent model introduced in the {\it Theory} section and discussed in details in \cite{SupMat}.  With this theoretical approach  we can implement the exact experimental field modulation, and  reproduce very well the line-splitting as seen in Fig.~\ref{Fig:VaryingAmplitude} (a). In addition, we can also obtain a good agreement between theory and experiments for the range of atom number probed in Fig~\ref{Fig:ModulationResponse} (b), with the scattering length as a single adjustable parameter. The result is shown in Fig.~\ref{Fig:ModulationResponse} (c). In this plot, the scissors mode frequency is shown as a dashed line, showing that the departure from linear response  causes a significant shift of the signal. Finally this allows us to conclude on the scattering length for which our observations at different atom numbers and $\theta_0=12^\circ$ are best reproduced: $a=68(5)\,a_0$, where the main contribution to the error is coming from the systematic uncertainty in the atom number \cite{SupMat}. When on the other hand we use only the low-amplitude data $\theta_0=5^\circ$ at $N=390$ shown in Fig.~\ref{Fig:VaryingAmplitude}, we obtain $a=67(6)\,a_0$, in agreement.\par
The non-linear coupling between the scissors and other low-lying modes provides us with a new tool to excite the latter. In the last part of this paper, we use this to study the properties of the lowest mode. The mode coupling arises at large angles between the field and the droplet. To probe this regime, we perform a $90^\circ$ rotation of the field at constant $B_0 = 800\,\mg$ in a time $t\simeq3\,\ms$. Systematic imaging errors prevent the direct observation of angle oscillations. Nevertheless we observe that the droplet quickly rotates by $90^\circ$. Via this field orientation quench, we obtain clear evidence for excitation of a collective mode, seen as a time-oscillation of the droplet length. These oscillations are strongly damped and we are able to observe them up to times of about $20\,\ms$ \cite{SupMat}. We infer that these oscillations correspond to an excitation of the lowest frequency collective mode of the system, observed in \cite{Chomaz:2016}. It consists essentially in a compression of the long axis of the droplet. Performing simulations of the equations of motion, we find that non-linearities must be taken into account. We therefore compare our experiments to numerical solutions of the equations of motion as above, applying our exact experimental field sequence.\par
\begin{figure}[hbtp]
\includegraphics[width=\columnwidth]{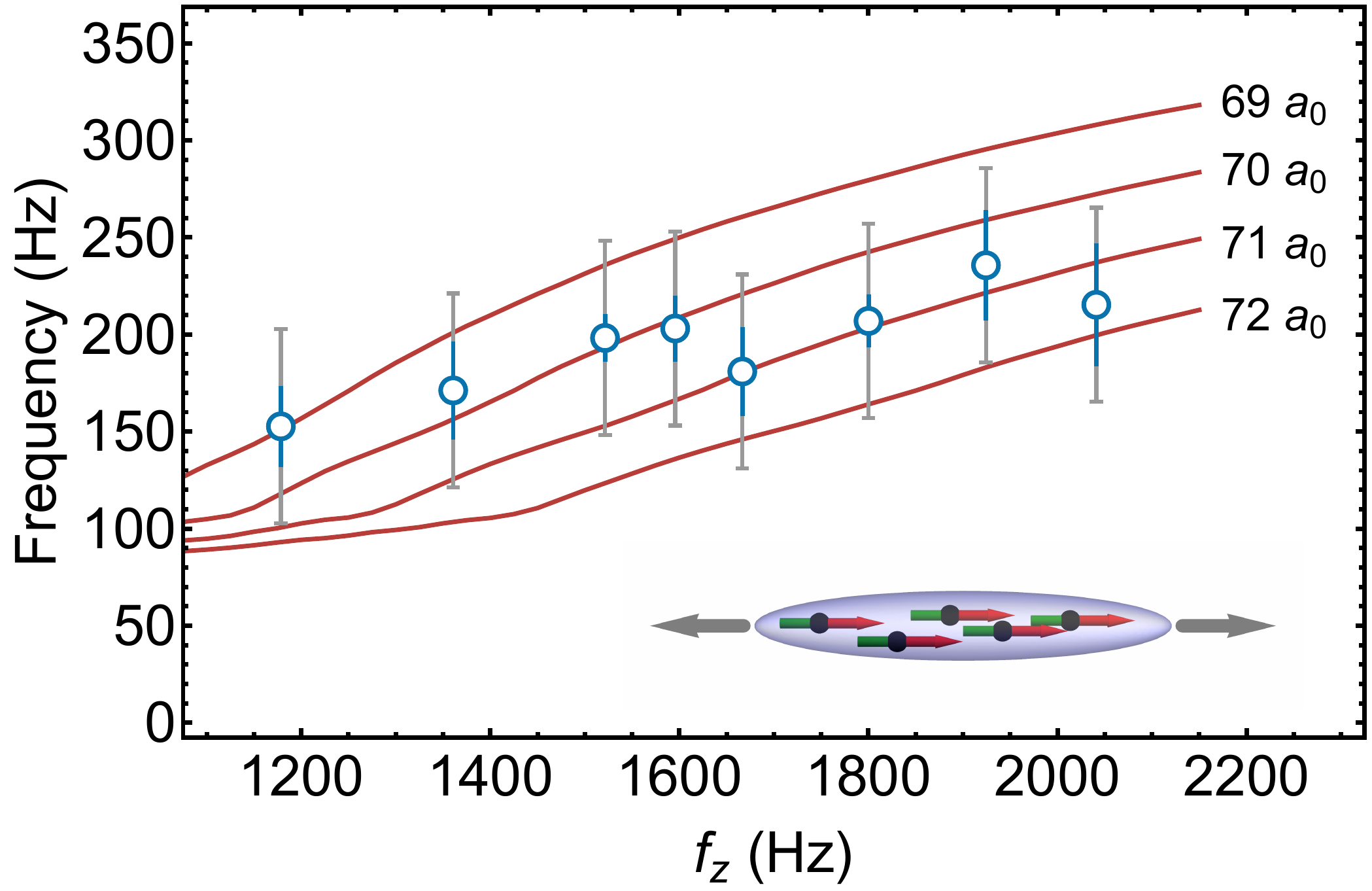}
\caption{Long axis oscillation frequency following a $90^\circ$ field angle quench. Experiments are performed with an atom number of $690(150)$. The error bar in blue shows the result of a squared sum of the fit standard error and the fitted decay time. The grey error bars show a $100\,\Hz$ interval corresponding to the short observation time. The red lines are theory calculations using the equations of motion for different scattering length (indicated on the figure).}
\label{Fig:Quadrupole}
\end{figure}

We vary the $z$ trapping frequency $f_z$, and record the variation in the observed frequency of oscillation over the first $10$ to $20\,\ms$. The experimental data is shown in Fig.~\ref{Fig:Quadrupole}, where we observe very little shift. We compare our measurements to theory and obtain relatively good agreement, though the increase in frequency predicted by theory is not clear in the experimental data. The best agreement is obtained for $a=70.5(6.0)\,a_0$. This value is compatible with the scattering length extracted from the scissors mode parametric excitation.\par
The scattering length values extracted here allow us to conclude on the interactions present in Dy ultracold samples at $B=800\,\mg$ so that $a=\abg$. The error-weighted mean experimental value from the two measurements is $\abg=69(4)\,a_0$. Values obtained in non-condensed samples \cite{Tang:2015,Maier:2015,Tang:2016} are consistently higher than that reported in quantum droplets \cite{Schmitt:2016}, though with large error bars. This might stem from the effect theoretically predicted in \cite{Oldziejewski:2016b} of an effective dipole-dipole interaction dependent on collisional energy. Systematic measurements, for instance using the lattice method of \cite{Chomaz:2016} could shed further light into the two-body interactions, which can no-longer be described by a simple addition of the DDI potential and the effective contact interaction potential \cite{Bortolotti:2006,Oldziejewski:2016a}. An interesting direction to further explore is to try to compare precise measurements of collective mode frequencies on one side and critical atom number in self-bound droplets on the other side. Indeed, these precise spectroscopic probes should hold a signature of departure from the eGPE description of the system, a breakdown of the local density approximation should be expected, and the critical atom number might depend on details of the potential beyond the $s$-wave scattering length as was shown for Bose mixtures in \cite{CikojeviC:2017}. Quantum Monte-Carlo works like refs.~\cite{Saito:2016,Macia:2016,Cinti:2017} will need to include two-body interaction potentials faithful to the real Dy-Dy potential.\par
In conclusion, our work sheds new light on dipolar quantum droplets, demonstrating a macroscopic collective mode shared with atomic nuclei. In both systems this mode is due to internal interactions, though for the present case of dipolar quantum droplets the rotational symmetry is broken by the external homogeneous magnetic field, while it is spontaneously broken in nuclei. We showed that the method developed here is a powerful probe for interactions in dipolar quantum droplets, and its systematic application should lead to the observation of physics beyond the current theoretical level. 
\begin{acknowledgments}
The Stuttgart group would like to acknowledge discussions with John Bohn. S.S. likes to thank Russell Bisset for useful comments and acknowledges funding from the Provincia Autonoma di Trento and the QUIC grant of the European  Horizon2020 FET program. This work is supported by the German Research Foundation (DFG) within FOR2247. I.F.B. and T.L. acknowledge support from the EU within Horizon2020 Marie Sk\l odowska Curie IF (Grants No. 703419 DipInQuantum and No. 746525 coolDips, respectively). T.L. acknowledges support from the Alexander von Humboldt Foundation through a Feodor Lynen Fellowship.
\end{acknowledgments}
\bibliography{Scissors}

\vspace{1.3cm}
\newpage
\onecolumngrid
\begin{center}
	{\Large \textbf{Supplemental Material}}
\end{center}
\twocolumngrid
\vspace{1cm}

\setcounter{figure}{0}   
\setcounter{equation}{0}   
\renewcommand{\thefigure}{S\arabic{figure}}
\renewcommand{\theequation}{S\arabic{equation}}

\vspace{.5cm}\textbf{Theory}\par
\textbf{Sum rules}
We calculate here the scissors mode frequency from Eqs.~(3)~and~(4) of the main text. Our results are general and can be applied to any known density distribution, we apply them here to a Gaussian ansatz:
\begin{equation}
	\psi(\boldsymbol{r})=\frac{\sqrt N}{\pi^{3/4}\bar \sigma^{3/2}}e^{\sum_\alpha -\frac{\alpha^2}{2\,\sigma_\alpha^2}}\label{Eq:psi}
\end{equation} 
with $\alpha \in \{x,y,z\}$ and $\bar\sigma^3=\sigma_x\sigma_y\sigma_z$. The density is given by $n(\boldsymbol r) = |\psi (\boldsymbol{r} )|^2$. Using the sum-rule formalism \cite{Pitaevskii:2016}, the scissors mode frequency is given by
 \begin{equation}
 \hbar \omega_{sc} = \sqrt{ \frac{m_1}{m_{-1}}},
 \label{w_sc}
 \end{equation}
where $m_1 = \hbar^2 \int d\omega \omega S_{L_z} (\omega)$ is the energy-weighted moment and $m_{-1} = \hbar^2 \int d\omega/\omega S_{L_z} (\omega)$ is the inverse energy-weighted moment. The dynamical structure factor $S_{L_z} (\omega)$ relative to the operator $\hat{L}_z = \sum_k \hat{x}_k \hat{p}_k^y - \hat{y}_k \hat{p}_y^x$ is
\begin{equation}
S_{L_z} (\omega) = Q^{-1} \sum_{m,n} e^{-\beta E_m} \left| \left< n |\hat{L}_z | m \right> \right|^2 \delta(\hbar \omega - \hbar \omega_{mn}),
\end{equation}
where $Q = \sum_m e^{-\beta E_m}$ is the partition function, $|n\rangle$ and $\omega_{mn} = (E_n - E_m)/\hbar$ are, respectively, the eigenstates and the transition frequencies of the Hamiltonian. 
Here, the Hamiltonian reads 
\begin{equation}
\begin{split}
\hat{H} &= \hat{H}_{\rm kin} + \hat{H}_{\rm ho} + \hat{H}_{\rm ci} + \hat{H}_{\rm dd} \\
& = \sum_{k=1}^{N} \frac{\hat{p}_k^2}{2m} + \sum_{k=1}^{N} \frac{m}{2} \left( \omega_{\perp}^2 (\hat{x}_k^2 + \hat{y}_k^2) + \omega_z^2 \hat{z}_k^2  \right) \\
& + g \sum_{i,k} \delta(\mathbf{\hat{r}}_i - \mathbf{\hat{r}}_k) \\
& + \frac{1}{2}\sum_{i,k} \frac{C_{\rm dd}}{4\pi |\mathbf{\hat{r}}_i - \mathbf{\hat{r}}_k |^3} \left[ 1 - \frac{3\left( \hat{y}_i - \hat{y}_k \right)^2}{|\mathbf{\hat{r}}_i - \mathbf{\hat{r}}_k |^2} \right],
\end{split}
\end{equation}
with $N$ the number of particles and $g$ the coupling constant of contact interactions. The last term holds for dipolar interactions, $C_{\rm dd}=\mu_0\mu^2$.

Using the Gaussian ansatz in Eq.~(4) of the main text, the inverse energy-weighted moment $m_{-1}$ can be rewritten
\begin{equation}
m_{-1} = \frac{1}{4} m N \frac{(\sigma_y^2 - \sigma_x^2)^2}{\sigma_x^2 + \sigma_y^2},
\label{mMinus1}
\end{equation}
where we used $\langle r_i^2 \rangle = \sigma_i^2/2$. The $m_1$ sum rule can be expressed in terms of commutators through the following expression \cite{Pitaevskii:2016}:
\begin{equation}
m_1 = \frac{1}{2} \left< [\hat{L}_z,[\hat{H},\hat{L}_z]] \right>.
\end{equation}
All the terms of the hamiltonian commute with $\hat{L}_z$ except for the dipolar interaction term which breaks the rotational symmetry. Using the commutation relation between $\hat{\alpha}$ and $\hat{p}_{\alpha}, \alpha \in \{x,y,z \}$, $[ \hat{\alpha},\hat{p}_{\alpha}  ] = i \hbar$, the commutator $[\hat{H},\hat{L}_z]$ is 
\begin{equation}
[\hat{H},\hat{L}_z] = - i \hbar \sum_{i,k} \frac{3 C_{\rm dd}}{4 \pi |\boldsymbol{\hat{r}}_i - \boldsymbol{\hat{r}}_k |^5  } \left( \hat{y}_i - \hat{y}_k \right) \left( \hat{x}_i - \hat{x}_k \right).
\end{equation}
Then, it is straightforward to calculate the second commutator, which yields
\begin{equation}
\begin{split}
& [\hat{L}_z,[\hat{H},\hat{L}_z]]  = \\
& \hbar^2 \sum_{i,k} \frac{3 C_{\rm dd}}{4 \pi |\boldsymbol{\hat{r}}_i - \boldsymbol{\hat{r}}_k |^5} \left[\left( \hat{y}_i - \hat{y}_k \right)^2 - \left( \hat{x}_i - \hat{x}_k \right)^2 \right].
\end{split}
\end{equation}
The $m_1$ sum rule can then be written as
\begin{equation}
m_1 = \frac{\hbar^2}{2} \left< W(\boldsymbol{r}, \boldsymbol{r'})  \right>,
\end{equation}
with
\begin{equation}
\begin{split}
W(\boldsymbol{r}, \boldsymbol{r'}) & = \frac{3 C_{\rm dd}}{4 \pi |\boldsymbol{r} - \boldsymbol{r'} |^5} \left[\left( y - y' \right)^2 - \left( x - x'\right)^2 \right] \\
& = V_{dd}^x(\boldsymbol{r}, \boldsymbol{r'}) - V_{dd}^y(\boldsymbol{r}, \boldsymbol{r'}),
\end{split}
\end{equation}
where $V_{dd}^{\alpha}(\boldsymbol{r}) = \frac{C_{\rm dd}}{4\pi r^3}\left( 1 - 3 \frac{\alpha^2}{r^2}\right), \alpha \in \{x, y, z\}$ and one can write
\begin{equation}
m_1 = \frac{\hbar^2}{2} \left( \langle V_{dd}^x\rangle - \langle V_{dd}^y\rangle \right),
\end{equation}
where $\left< V_{dd}^{\alpha} \right > = \int d \boldsymbol{r} d\boldsymbol{r'} n(\boldsymbol{r}) V_{dd}^{\alpha}(\boldsymbol{r} - \boldsymbol{r'}) n(\boldsymbol{r'})$. Using the Gaussian ansatz, the expression of $\langle V_{\rm dd}^\alpha\rangle$ is a known result \cite{Glaum:2007} and the moment $m_1$ finally reads
\begin{equation}
m_1 = \frac{\hbar^4 N^2 a}{m \overline{\sigma}^3 \sqrt{2 \pi}} \varepsilon_{dd} \left(f_{\rm dip}\left(\frac{\sigma_x}{\sigma_y}, \frac{\sigma_z}{\sigma_y} \right) - f_{\rm dip}\left(\frac{\sigma_y}{\sigma_x}, \frac{\sigma_z}{\sigma_x} \right) \right),
\label{m1}
\end{equation} where $f_{\rm dip}(x,y)$ is defined in \cite{Giovanazzi:2006}. By dividing Eq.~(\ref{m1}) by Eq.~(\ref{mMinus1}) one immediately gets the scissors mode frequency
\begin{equation}
\resizebox{.9\columnwidth}{!} 
{$
\omega_{\rm sc}^2 =  \frac{4 \hbar^2 N a_{\rm dd}}{m^2 \sqrt{2 \pi} \overline{\sigma}^3}\frac{\sigma_x^2 + \sigma_y^2}{(\sigma_y^2 - \sigma_x^2)^2} \left( f_{\rm dip}\left(\frac{\sigma_x}{\sigma_y},\frac{\sigma_z}{\sigma_y}\right) - f_{\rm dip}\left(\frac{\sigma_y}{\sigma_x},\frac{\sigma_z}{\sigma_x}\right)\right)$.}\label{Eq:Frequency}
\end{equation}

\vspace{.5cm}\textbf{Equations of motion}\par
To obtain the equations of motion, we use the method of \cite{Yi:2001} employing the Lagrangian formalism, with the following ansatz for the droplet wavefunction:
\begin{equation}
	\psi(\boldsymbol{r})=\frac{\sqrt N}{\pi^{3/4}\bar \sigma^{3/2}}e^{\sum_k \left(-\frac{r_k'^2}{2\,\sigma_k^2}+i\beta_k r_k'^2\right)+ix'y'\Omega}\label{Eq:psi}\\
\end{equation}
with $\bar\sigma^3=\sigma_x\sigma_y\sigma_z$ and:
\begin{eqnarray}	\left\{\begin{matrix}
	x'&=&\cos\theta\,x-\sin\theta\,y\\
	y'&=&\sin\theta\,x+\cos\theta\,y\\
	z'&=&z
	\end{matrix} \right.
\end{eqnarray}
This ansatz is the usual Gaussian variational ansatz with an additional angular degree of freedom $\theta$. We calculate the Lagrangian as in \cite{Yi:2001}. The Euler-Lagrange (E-L) equations for the parameters $\beta_k$ and $\Omega$ read: 
\begin{eqnarray}
	\beta_k&=&\frac{m}{2\hbar\,\sigma_k}\frac{d\sigma_k}{dt}\\
	\Omega&=&-\frac m\hbar\,\frac{\sigma_x^2-\sigma_y^2}{\sigma_x^2+\sigma_y^2}\,\frac{d\theta}{dt}.
\end{eqnarray}
The equations of motion then derive from the E-L equations for the parameters $\sigma_i$ and $\theta$. The linearization of these equations allow to calculate the collective modes of the system, in full agreement with the scissors mode expression in the main text and the collective modes found in \cite{Wachtler:2016}.\par
\vspace{.5cm}\textbf{Self-bound quantum droplets, trap-bound quantum droplets and solitons}\par
In the geometry considered here of a tight confinement perpendicular to the magnetic field, it was shown in previous theory works at the mean-field level that soliton solutions exist \cite{Tikhonenkov:2008,Koberle:2012}. We differentiate here self-bound quantum droplets, quantum droplets and solitons using the following classification (for all three cases, the mean-field interaction is attractive, as opposed to a BEC solution):
\begin{itemize}
	\item \textit{Self-bound quantum droplets} exist thanks to the interplay between mean-field (MF) attraction and beyond-mean-field repulsion (BMF), and remain bound without the presence of a trap. 
	\item \textit{Trap-bound quantum droplets} also exist thanks to the interplay between MF attraction and BMF repulsion, but necessitate an external trapping to exist and compensate kinetic energy. 
	\item \textit{Solitons} exist at the mean-field level, where even neglecting the BMF term, kinetic energy and mean-field attraction compensate each other. 
\end{itemize}
\begin{figure}[hbtp]
\includegraphics[width=\columnwidth]{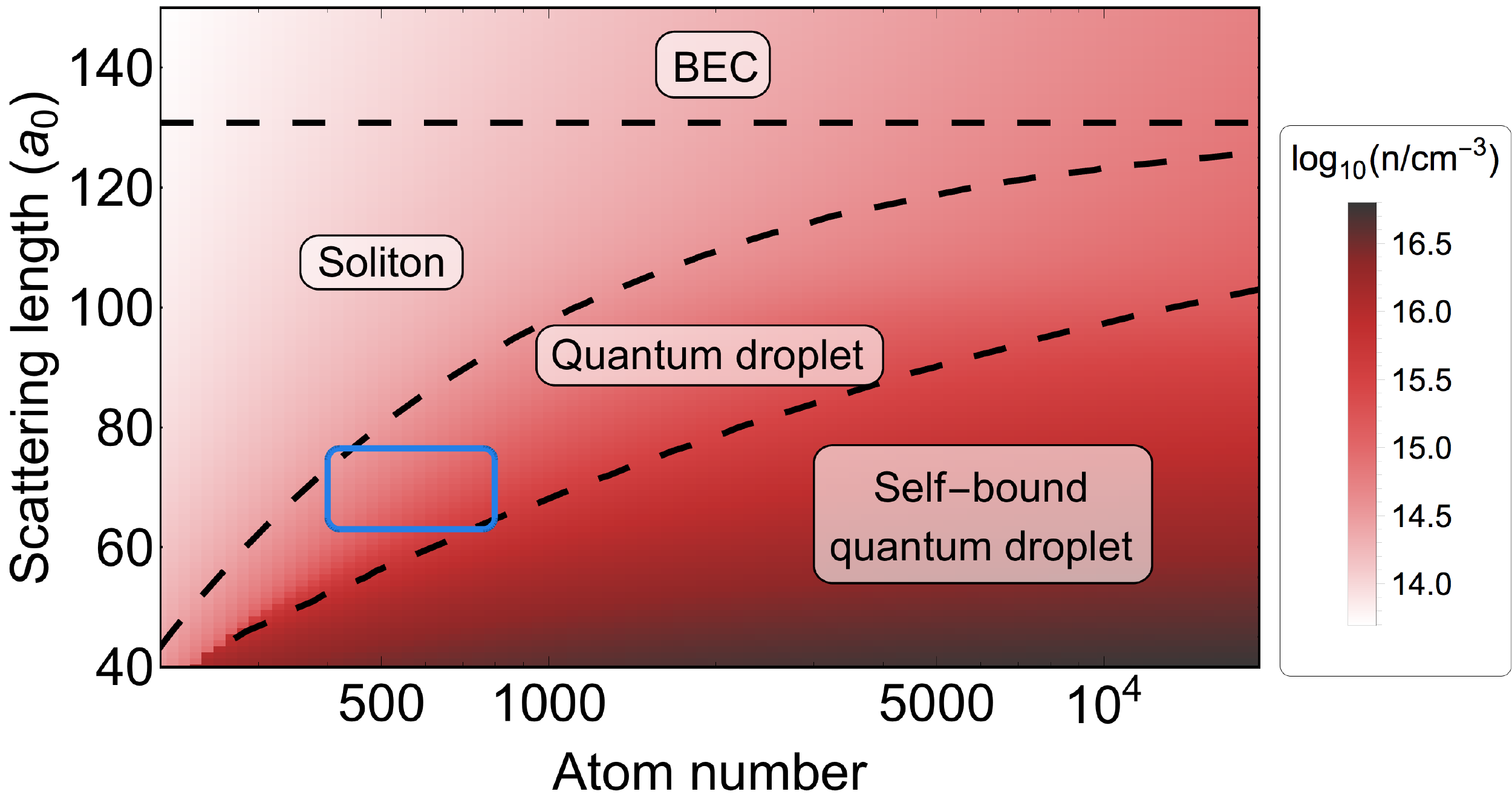}
\caption{Phase diagram showing the different solutions discussed in the text as a function of atom number and scattering length. The blue rectangle shows the atom number range probed in our experiments, and the uncertainty in scattering length, showing that we work in the quantum droplet regime where solitons are unstable but trapping is necessary to bind the quantum droplet. The coloring shows density in log scale.}
\label{Fig:DropletsAndSolitons}
\end{figure}

We represent in Fig.~\ref{Fig:DropletsAndSolitons} a phase diagram obtained in the Gaussian ansatz. We show the boundaries of existence of a self-bound, trap-bound quantum droplet, soliton and standard BEC in the trap used in the experiments ($f_x=f_y=40\,\Hz$, $f_z=950\,\Hz$), as a function of atom number and scattering length. Our experiments take place in a region where a soliton is unstable but trapping is necessary to bind the droplet. The scissors mode does not vanish in the absence of trapping, and the self-bound character of dipolar quantum droplets was established experimentally in \cite{Schmitt:2016}. However, self-bound quantum droplets are very short-lived rendering challenging the excitation of this mode. Furthermore, the observation of self-bound droplets requires a homogeneous levitation force usually obtained via a magnetic field gradient, this is very challenging to combine with a modulated field orientation. For such reasons we performed our experiments in an optical dipole trap.\par
\vspace{.5cm}\textbf{Single droplet creation}\par
To create single droplets in the $xy$-plane, we use our experimental sequence developed in \cite{Wenzel:2017}. We start with a BEC in a pancake-like trap with the field along $\hat z$, we then tilt the field in the plane (along $\hat y$), the in-plane trap anisotropy ($f_x/f_y$) is initially adjusted so as to obtain a single droplet (see \cite{Wenzel:2017}). Once the field is along $\hat y$ we re-shape the trap so as to obtain the cylindrically-symmetric geometry given in the main text ($f_x=f_y=40(1)\,\Hz$, $f_z=950\,\Hz$).\par 
We image these droplets using phase-contrast imaging, the resolution of the imaging system is $1\,\mum$ according to Rayleigh's criterion, this corresponds to an in-situ Gaussian size ($\sigma_x\approx0.6\,\mum$), at rest, the droplets are smaller than this along $\hat x$. \par
\vspace{.5cm}\textbf{Atom number uncertainty and non-condensed atoms}\par
Little is known about the full in-trap density distribution at finite temperature including the thermal atoms and the quantum droplet.  In order to extract the atom number within the droplet and estimate the systematic uncertainty, we perform two different counting procedures. First, we apply a dual Gaussian fit to the whole image: As a simple sum of two density distributions, this neglects the possible interaction between thermal and condensed atoms and typically underestimates the condensed atom number. Second, we fit a single Gaussian to the cloud center assuming a local density background set by the thermal cloud. This does not estimate that the thermal density in the droplet region is zero, but rather that it is flat. The two methods give on average a $25\,\%$ difference. We take this to be the main source of systematic uncertainty in the number determination. The quoted numbers are the ones obtained by the single fit. The number as a function of time is represented in Fig.~\ref{Fig:NVsT}.\par
Another source of uncertainty comes from the possibility that non-condensed atoms are trapped within the droplet by the attractive potential created by the condensed atoms. To estimate the number of atoms possibly trapped, we consider the mean-field potential created by the droplet as a trapping potential. Solving Schr\"odinger's equation numerically gives a good estimate of the number of states and their energy splitting within this trap. We typically find $2-7$ states. Assuming the ground state is the condensed state thus fixing the chemical potential, we can estimate the population in excited states with a Bose-Einstein distribution. This results in all our parameter range to less than 10 non-condensed atoms trapped within the droplet, a negligible quantity. We calculate that the population in the harmonic oscillator states of the trapping potential should be on the order of a few hundred, in agreement with the observed thermal atom numbers outside the droplet.\par
\begin{figure}[hbtp]
\includegraphics[width=.8\columnwidth]{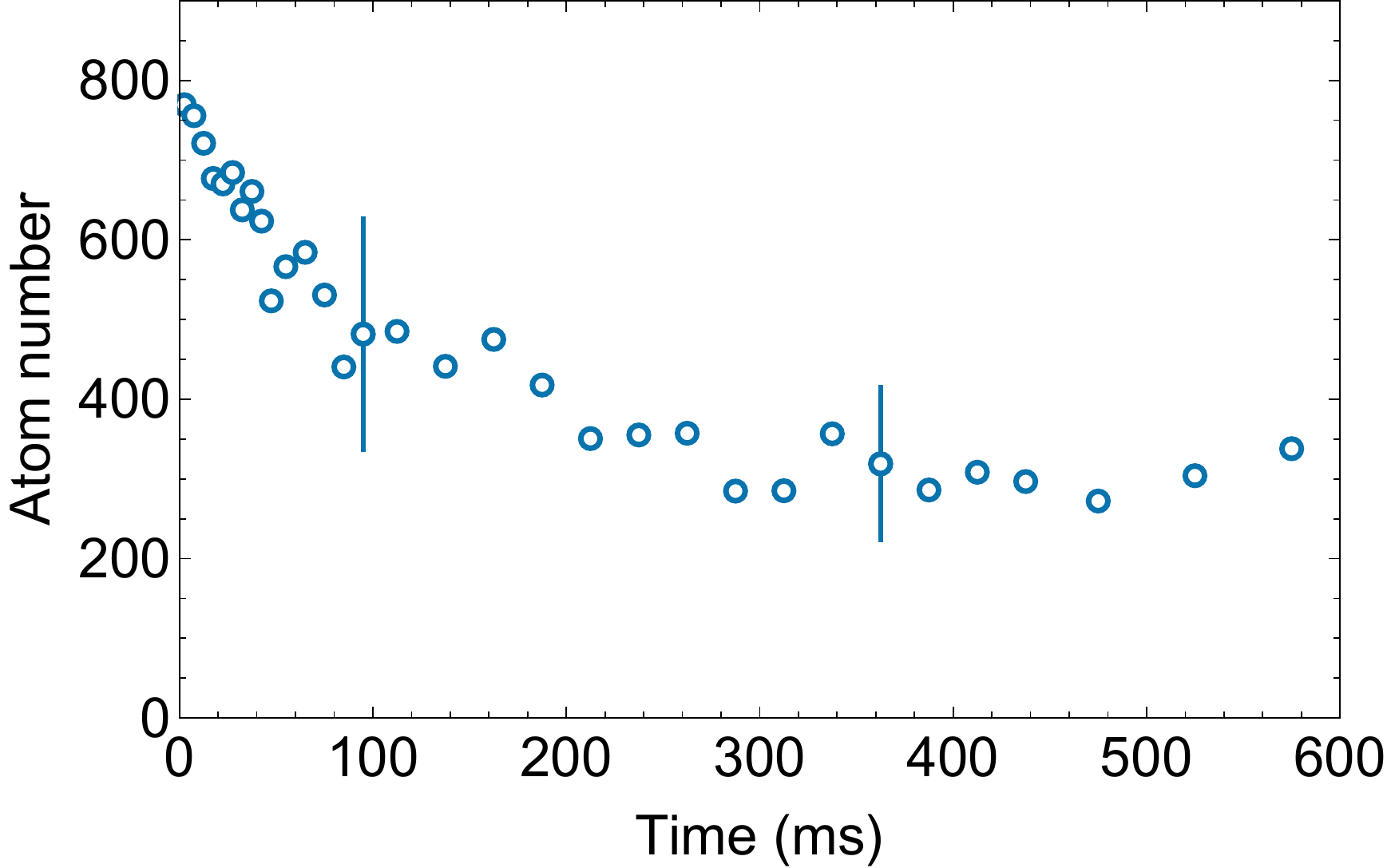}
\caption{Atom number as a function of time, the indicative error bar shows the uncertainty on the atom number, dominated by systematic effect described in the text.}
\label{Fig:NVsT}
\end{figure}
\vspace{.5cm}\textbf{Quadrupole oscillations}\par
Following the quench of the magnetic field orientation of $90^\circ$, we record the length of the droplet as a function of time. For this we need to fit the angle of the droplet. Our precision on the measurement of the direction of the droplet as well as that of the field does not allow to conclude positively on the observation of clear oscillations in angle. The length however is clearly modulated with time, decaying over times of the order of $20\,\ms$, this is shown in Fig.~\ref{Fig:LengthOsc}, the frequency obtained from the fits shown in dashed lines is plotted in the main text.\par
\begin{figure}[hbtp]
\includegraphics[width=.8\columnwidth]{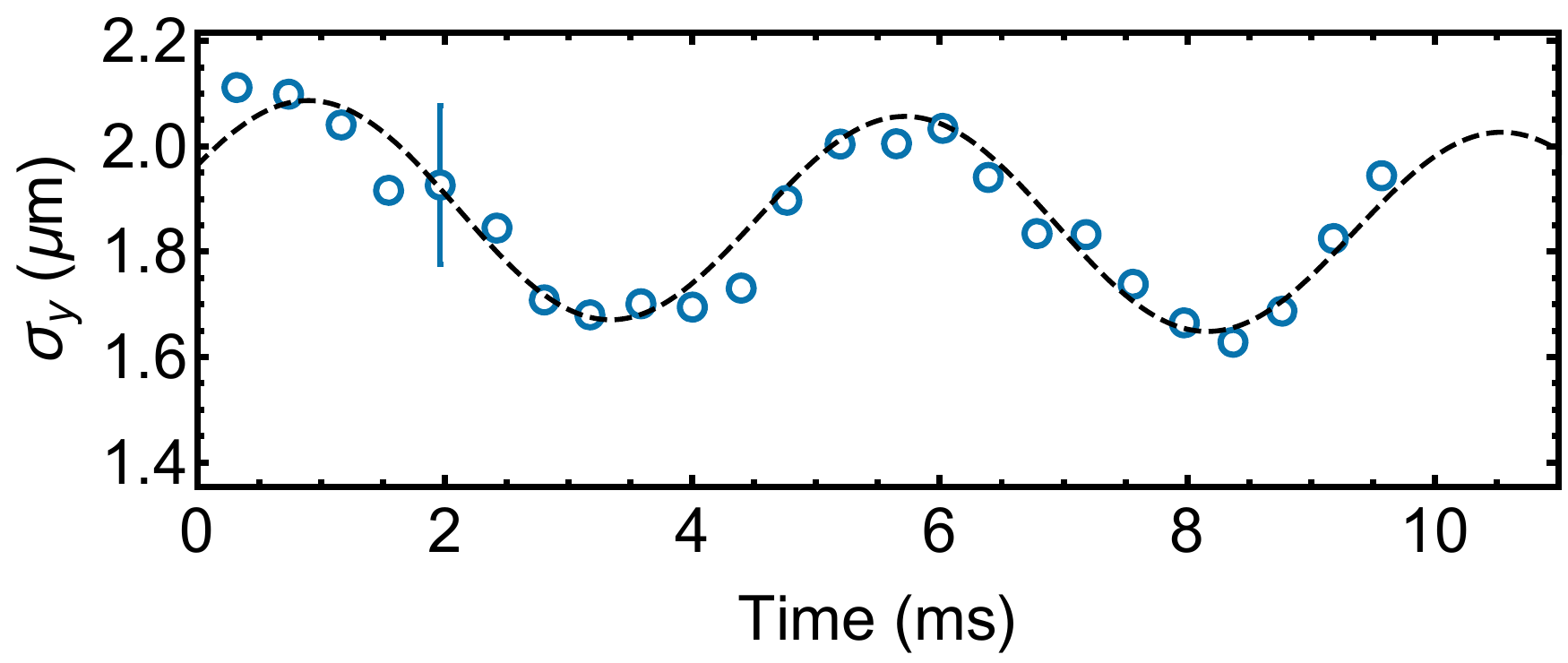}
\includegraphics[width=.8\columnwidth]{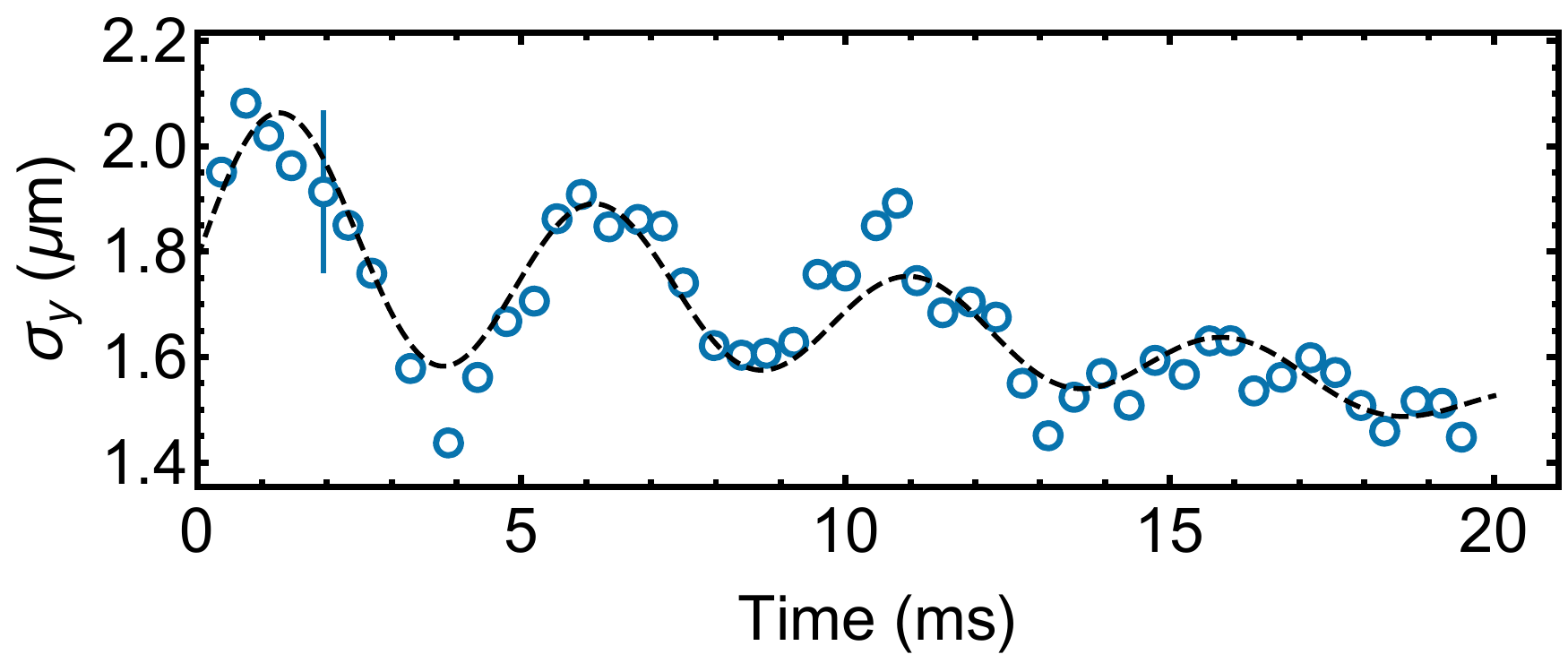}
\caption{Time evolution of the long droplet size as a function of time following a $90^\circ$ angle quench. For two different $z$ trapping frequencies. Each point is the mean of 4 measurements, an error bar indicates the typical standard error. Data are fitted with phenomenological damped sinusoids (with a drift) to extract frequency, grey dashed lines.}
\label{Fig:LengthOsc}
\end{figure}

\begin{figure}[hbtp]
\includegraphics[width=\columnwidth]{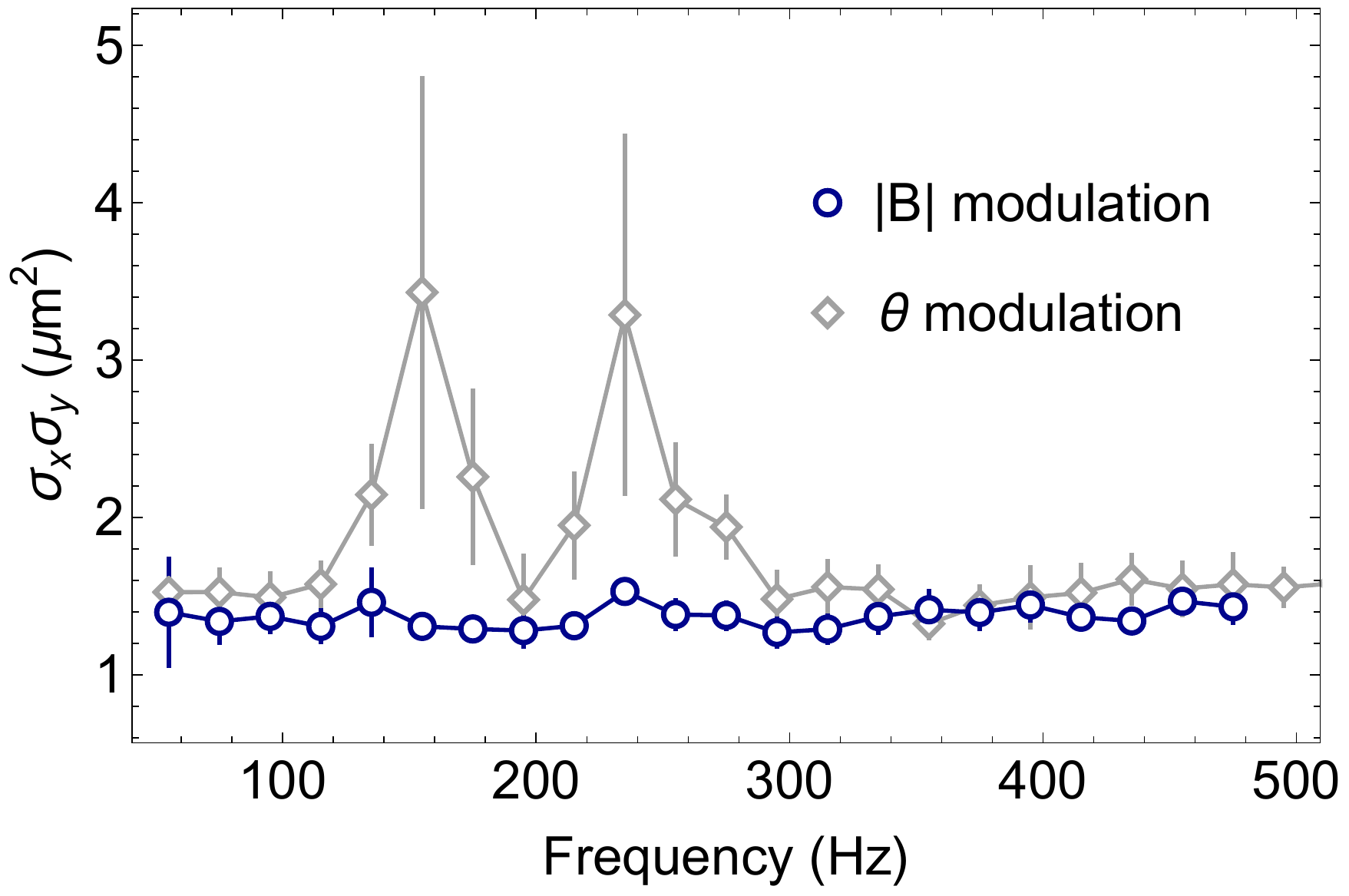}
\caption{Response to a modulation of the magnitude of the homogeneous magnetic field at constant orientation (circles), compared to a modulation of the angle of the field (diamonds). The absence of response when modulating the field magnitude indicates clearly that the response when modulating the angle comes only from the modulation of the direction of the field.}
\label{Fig:MagnitudeModulation}
\end{figure}

\vspace{.5cm}\textbf{Modulation of the magnetic field magnitude with constant orientation}\par
To verify that the observed response (see Fig.~(2) of main text) stems only from the modulation of the field orientation, we have modulated the magnetic field magnitude at fixed orientation. The magnitude is varied with amplitude $\delta B=80\,\mg$ so that $\delta B/B_0=0.1$, greater than the magnitude modulation occurring with the orientation modulation $\delta B/B_0\leqslant0.04$. The data is plotted in Fig.~\ref{Fig:MagnitudeModulation}. We observe no response, despite a greater magnitude modulation. Therefore we conclude that the response observed in the text is due to the modulation in the field orientation.\par

\begin{figure}[hbtp]
\includegraphics[width=\columnwidth]{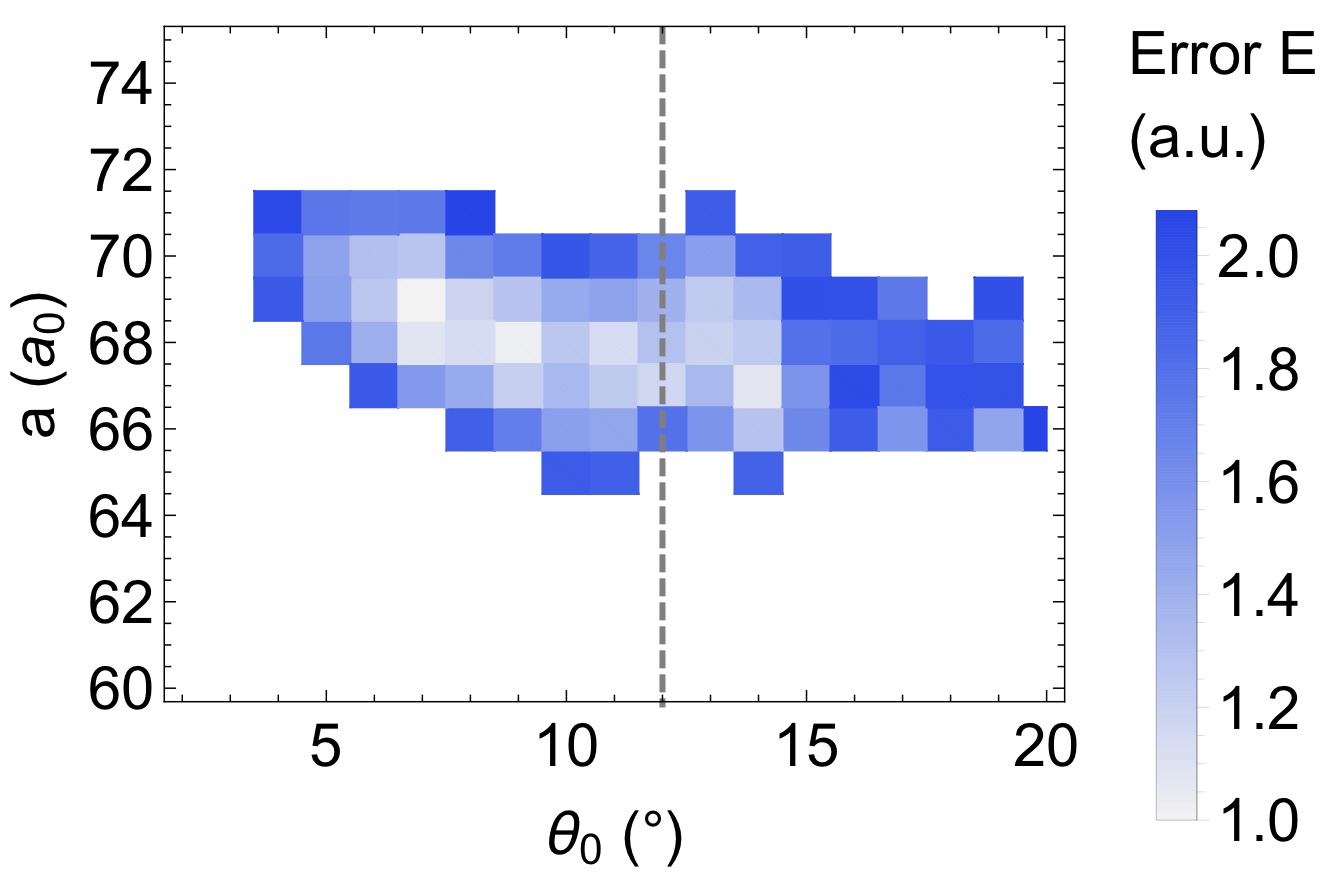}
\caption{Error $E$ (see text) used to find the best fit parameter $a$. One observes that $E$ is minimized for modulation amplitudes close to the experimental one ($\theta_0=12^\circ$, dashed line), which confirms that the lineshape characteristics arise from this large amplitude.}
\label{Fig:errorPlot}
\end{figure}

\begin{figure}[hbtp]
\includegraphics[width=.75\columnwidth]{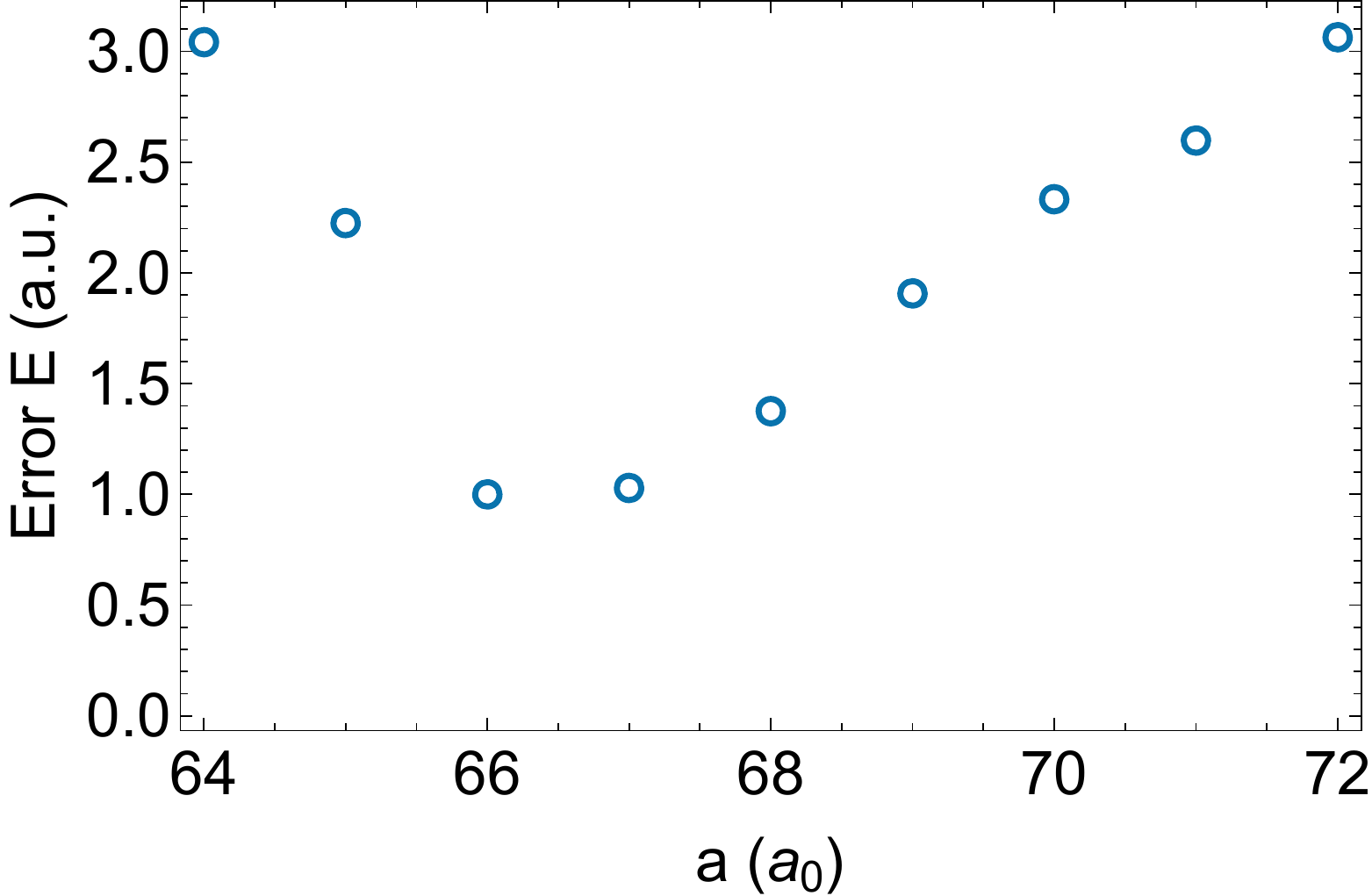}
\caption{Error $E$ (see text) used to find the best fit parameter $a$ using only the $\theta_0=5^\circ$ data. The data is compared to the model predictions fixing also $\theta_0=5^\circ$. }
\label{Fig:errorPlot5deg}
\end{figure}

\vspace{.5cm}\textbf{Systematic error on scattering length}\par
By comparing the experimental data presented in the main text with solutions of the equations of motion outlined above, we extract a value for the scattering length. The equations of motion are solved applying the field angle modulation as in the experiment. The amplitude is varied following the exact dependence that is used experimentally ($\theta_{\rm m}=\theta_0/\sqrt{f/\SI{100}{\hertz}}$), with $\theta_0=12^\circ$. We vary the scattering length to find the best fit with the experimental data. To do so we minimize an error $E=\sum_i(d_i-t_i)^2$ with $d_i$ the $\sigma^2$ data points shown in main text and $t_i$ the corresponding theory prediction. To account for a possible miscalibration of the imaging, or other sources of error on the absolute size of the droplet versus time, we rescale the data and theory to the same minimum and maximum value. This way we find the scattering length for which the lineshape is best reproduced. The absolute response $\sigma_x\sigma_y$ obtained for the best fit parameter $a$ is represented in Fig.~1 (c) from main text, it differs by about $30\,\%$ between theory and experiment. This difference could easily be explained by slight miscalibration of imaging and the simplifying assumptions of the model, which in particular does not allow any dissipation.\par

\begin{figure}[tb]
\includegraphics[width=\columnwidth]{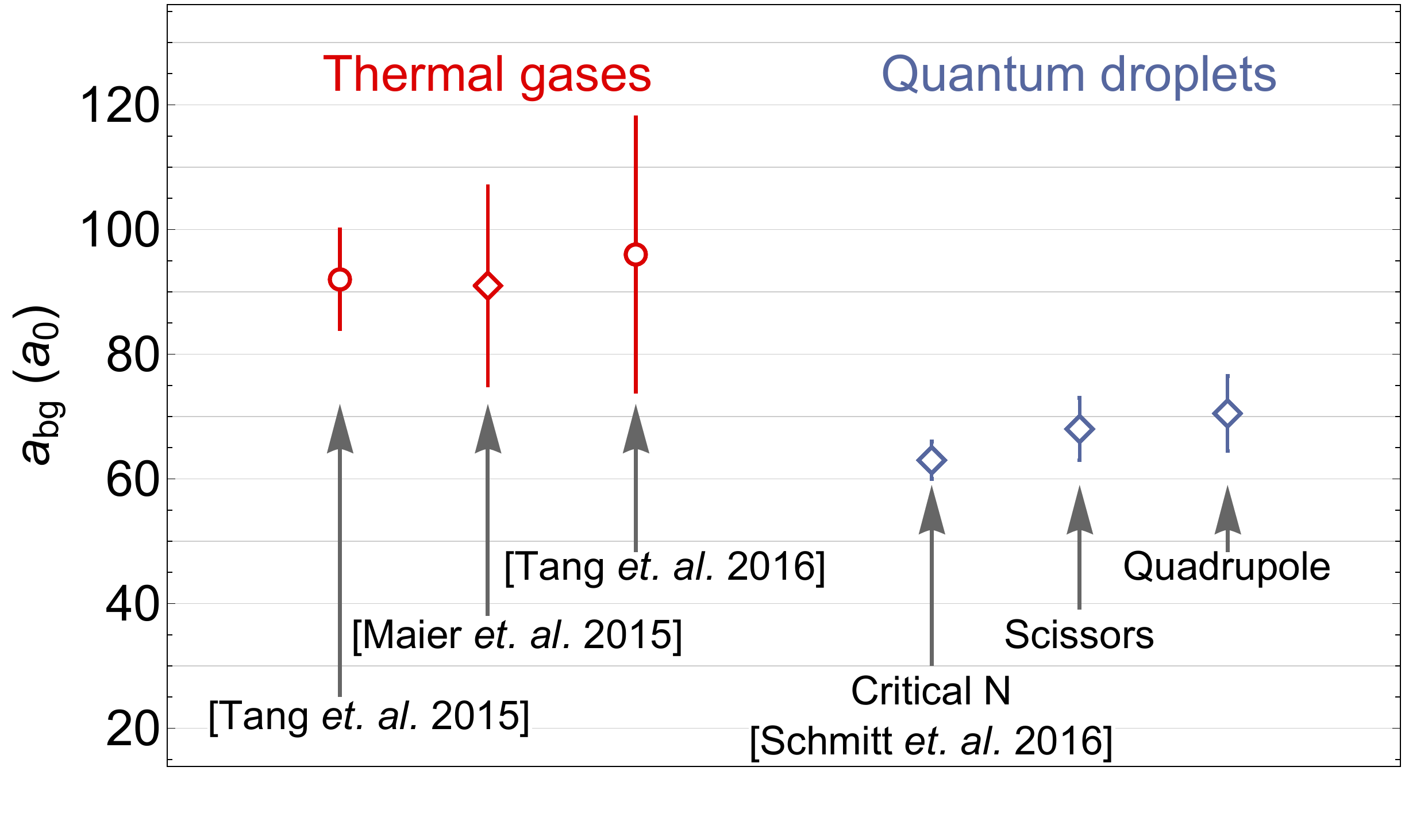}
\caption{Different reported values for the background scattering length of \textsuperscript{164}Dy: Observations on thermal gases by the Stanford group (Tang \textit{et. al.}~2015 \cite{Tang:2015} and Tang \textit{et. al.}~2016 \cite{Tang:2016}). Feshbach molecule spectroscopy by our group (Maier \textit{et. al.}~2015 \cite{Maier:2015}), and critical atom number in quantum droplets (Schmitt \textit{et. al.}~2016 \cite{Schmitt:2016}). The two last points show results from the present paper.}
\label{Fig:ScatteringLengthMeasurements}
\end{figure}

The lowest $E$ is obtained for $a=68\,a_0$. This error $E$ is doubled when varying $a$ by $3\,a_0$, which we take as a statistical error (see Fig.~\ref{Fig:errorPlot}). To this error we must add the systematic uncertainty due to that of the experimental atom number ($25\,\%$, given above). To do so, we rescale the experimental atom number by $\pm25\,\%$ and compare again with theory. We find that the lowest $E$ is then found for $64\,a_0$ when rescaling the atom number down, and $72\,a_0$ when rescaling it up. As a consequence we infer that the systematic error due to miscounting the atom number is $5\,a_0$. Other sources of uncertainty such as error on the actual droplet length, trapping frequency uncertainty are extremely weak and contribute negligibly with respect to the systematic errors above. We thus infer a scattering length measurement and uncertainty of $a=68(5)\,a_0$.\par 
We also extract the scattering length from the low amplitude data ($\theta_0=5^\circ$) data at $N=390$ reported in Fig.~2 of main text. We apply the same procedure as for the $\theta_0=12^\circ$ data.  The error versus theoretical scattering length is reported in Fig.~\ref{Fig:errorPlot5deg}, where we see that the error is minimized for $a\approx68\,a_0$. This data is less prone to error due to strong departure from linear response not taken into account in our model, but due to the lower signal we obtain a slightly larger error, resulting in $a=67(6)\,a_0$.\par  
The same systematic sources of error apply to the scattering length obtained from the quadrupole oscillation frequency. The experimental standard error intervals extracted from the fits like the ones shown in Fig.~\ref{Fig:LengthOsc} are displayed in the main text Fig.~3. The mean atom number in these measurements is $N=690$ (with a standard deviation $95$). The theory for different scattering length is also shown. This results in an error due to frequency uncertainty of $2\,a_0$ and due to the systematic $N$ uncertainty of $5\,a_0$ and thus a squared sum of $5\,a_0$, resulting in the measurement $a=70.5(5)a_0$.\par
We summarize all background scattering length measurements for \textsuperscript{164}Dy published so far, together with the two values obtained in the present work in Fig.~\ref{Fig:ScatteringLengthMeasurements}. It includes the values obtained in thermal gases by the Stanford group: \cite{Tang:2015} and \cite{Tang:2016}, the estimate from molecular association spectroscopy on a Feshbach resonance at $76\,\g$ by our group \cite{Maier:2015} and finally our comparison of the self-bound quantum droplets critical atom number data to the eGPE in \cite{Schmitt:2016}.\par

\end{document}